\documentclass[superscriptaddress,prl,letterpaper,twocolumn]{revtex4}
\usepackage{amsfonts}
\usepackage{amsmath}
\usepackage{amssymb}
\usepackage{graphics}

\input{tcilatex}
\begin{document}

\title{The effect of random coupling coefficients on decoherence}
\author{Mario Castagnino}
\affiliation{CONICET-IAFE-IFIR-Universidad de Buenos Aires}
\author{Sebastian Fortin}
\affiliation{CONICET-IAFE-Universidad de Buenos Aires}
\author{Olimpia Lombardi}
\affiliation{CONICET-Universidad de Buenos Aires}

\begin{abstract}
The aim of this letter is to analyze the effect on decoherence of the
randomness of the coupling coefficients involved in the interaction
Hamiltonian. By studying the spin-bath model with computer simulations, we
show that such randomness greatly improves the \textquotedblleft
efficiency\textquotedblright\ of decoherence and, then, its physical meaning
deserves to be considered.
\end{abstract}

\maketitle

\paragraph{\textbf{Introduction.}}

Environment induced decoherence (EID) is usually tested by means of computer
simulations on models whose coefficients are, in general, taken as random
with the idea of simulating a generic situation (see, as examples, \cite%
{Zurek-1982}, \cite{Max}, \cite{1}, \cite{2}, \cite{3}, \cite{4}). This
widely spread strategy is reasonable for the coefficients of the particles'
states, but it is not natural for the coupling coefficients involved in the
interaction Hamiltonian: there is no reason to suppose that particles of the
same nature interact with each other with different strengths. The point
would be irrelevant if the randomness of the coupling coefficients did not
affect the overall phenomenon. The aim of this letter is to show that this
is not the case: the randomness of the coupling coefficients greatly
improves the \textquotedblleft efficiency\textquotedblright\ of decoherence.
This result has to be taken as a warning against the uncritical use of
random coupling coefficients for drawing conclusions about decoherence. We
will argue for this claim by means of the analysis of a well-known model.

\paragraph{\textbf{The spin-bath model.}}

This is a very simple model that has been exactly solved in previous papers
(see \cite{Zurek-1982}). We will study it from the general theoretical
framework for decoherence presented in a previous work \cite{CQG-CFLL-08}.
Let us consider a closed system $U=S+E$ where (i) the system $S$ is a
spin-1/2 particle $P$ represented in the Hilbert space $\mathcal{H}_{S}$,
and (ii) the environment $E$ is composed of $N$ spin-1/2 particles $P_{i}$,
each one represented in its own Hilbert space $\mathcal{H}_{i}$. The
complete Hilbert space of the composite system $U$\ is $\mathcal{H}=\mathcal{%
H}_{S}\otimes \left( \bigotimes\limits_{i=1}^{N}\mathcal{H}_{i}\right) $. In
the particle $P$, the two eigenstates of the spin operator $S_{S,%
\overrightarrow{v}}$\ in direction $\overrightarrow{v}$ are $\left\vert
\Uparrow \right\rangle $ and $\left\vert \Downarrow \right\rangle $, such
that $S_{S,\overrightarrow{v}}\left\vert \Uparrow \right\rangle =\frac{1}{2}%
\left\vert \Uparrow \right\rangle $\ and\ $S_{S,\overrightarrow{v}%
}\left\vert \Downarrow \right\rangle =-\frac{1}{2}\left\vert \Downarrow
\right\rangle $. In each particle $P_{i}$, the two eigenstates of the
corresponding spin operator $S_{i,\overrightarrow{v}}$\ in direction $%
\overrightarrow{v}$ are $\left\vert \uparrow _{i}\right\rangle $ and $%
\left\vert \downarrow _{i}\right\rangle $, such that $S_{i,\overrightarrow{v}%
}\left\vert \uparrow _{i}\right\rangle =\frac{1}{2}\left\vert \uparrow
_{i}\right\rangle $ and\ $S_{i,\overrightarrow{v}}\left\vert \downarrow
_{i}\right\rangle =-\frac{1}{2}\left\vert \downarrow _{i}\right\rangle $.
Therefore, a pure initial state of $U$ reads%
\begin{equation}
|\psi _{0}\rangle =(a\left\vert \Uparrow \right\rangle +b\left\vert
\Downarrow \right\rangle )\otimes \left( \bigotimes_{i=1}^{N}(\alpha
_{i}|\uparrow _{i}\rangle +\beta _{i}|\downarrow _{i}\rangle )\right)
\label{1}
\end{equation}%
where the coefficients $a$, $b$, $\alpha _{i}$, $\beta _{i}$ are such that
satisfy $\left\vert a\right\vert ^{2}+\left\vert b\right\vert ^{2}=1$ and $%
\left\vert \alpha _{i}\right\vert ^{2}+\left\vert \beta _{i}\right\vert
^{2}=1$. The self-Hamiltonians $H_{S}$ and $H_{E}$ of $S$ and $E$,
respectively, are taken to be zero. Then the total Hamiltonian $%
H=H_{S}+H_{E}+H_{SE}$ of the composite system $U$ results (see \cite%
{Zurek-1982}, \cite{Max})%
\begin{equation}
H=H_{SE}=S_{S,\overrightarrow{v}}\otimes \sum_{i=1}^{N}2g_{i}S_{i,%
\overrightarrow{v}}\otimes \left( \bigotimes_{j\neq i}^{N}I_{j}\right)
\label{2}
\end{equation}%
where $I_{j}$ is the identity operator on the subspace $\mathcal{H}_{j}$, $%
S_{S,\overrightarrow{v}}=\frac{1}{2}\left( \left\vert \Uparrow \right\rangle
\left\langle \Uparrow \right\vert -\left\vert \Downarrow \right\rangle
\left\langle \Downarrow \right\vert \right) $ and $S_{i,\overrightarrow{v}}=%
\frac{1}{2}\left( \left\vert \uparrow _{i}\right\rangle \left\langle
\uparrow _{i}\right\vert -\left\vert \downarrow _{i}\right\rangle
\left\langle \downarrow _{i}\right\vert \right) $. Under the action of $%
H=H_{SE}$, the state $|\psi _{0}\rangle $ evolves as $\left\vert \psi
(t)\right\rangle =a\left\vert \Uparrow \right\rangle |\mathcal{E}_{\Uparrow
}(t)\rangle +b\left\vert \Downarrow \right\rangle |\mathcal{E}_{\Downarrow
}(t)\rangle $ where $\left\vert \mathcal{E}_{\Uparrow }(t)\right\rangle
=\left\vert \mathcal{E}_{\Downarrow }(-t)\right\rangle $ and%
\begin{equation}
\left\vert \mathcal{E}_{\Uparrow }(t)\right\rangle
=\bigotimes_{i=1}^{N}\left( \alpha _{i}\,e^{-ig_{i}t/2}\,\left\vert \uparrow
_{i}\right\rangle +\beta _{i}\,e^{ig_{i}t/2}\,\left\vert \downarrow
_{i}\right\rangle \right)  \label{3}
\end{equation}

An observable of $U$, $O\in \mathcal{O}=\mathcal{H\otimes H}$, reads%
\begin{equation}
O=\left( 
\begin{array}{c}
s_{\Uparrow \Uparrow }\left\vert \Uparrow \right\rangle \left\langle
\Uparrow \right\vert \\ 
+s_{\Uparrow \Downarrow }\left\vert \Uparrow \right\rangle \left\langle
\Downarrow \right\vert \\ 
+s_{\Downarrow \Uparrow }\left\vert \Downarrow \right\rangle \left\langle
\Uparrow \right\vert \\ 
+s_{\Downarrow \Downarrow }\left\vert \Downarrow \right\rangle \left\langle
\Downarrow \right\vert%
\end{array}%
\right) \otimes \left( \bigotimes_{i=1}^{N}\left( 
\begin{array}{c}
\epsilon _{\uparrow \uparrow }^{(i)}\left\vert \uparrow _{i}\right\rangle
\left\langle \uparrow _{i}\right\vert \\ 
+\epsilon _{\downarrow \downarrow }^{(i)}\left\vert \downarrow
_{i}\right\rangle \left\langle \downarrow _{i}\right\vert \\ 
+\epsilon _{\downarrow \uparrow }^{(i)}\left\vert \downarrow
_{i}\right\rangle \left\langle \uparrow _{i}\right\vert \\ 
+\epsilon _{\uparrow \downarrow }^{(i)}\left\vert \uparrow _{i}\right\rangle
\left\langle \downarrow _{i}\right\vert%
\end{array}%
\right) \right)  \label{4}
\end{equation}%
In the typical situation studied by the EID approach, the system of interest 
$S$ is simply the particle $P$. Therefore, the relevant observables $O_{R}$
are obtained from eq.(\ref{4}) by making $\epsilon _{\uparrow \uparrow
}^{(i)}=\epsilon _{\downarrow \downarrow }^{(i)}=1$ and $\epsilon _{\uparrow
\downarrow }^{(i)}=0$:%
\begin{equation}
O_{R}=\left( \sum_{s,s^{\prime }=\Uparrow ,\Downarrow }s_{ss^{\prime
}}|s\rangle \langle s^{\prime }|\right) \otimes \left(
\bigotimes_{i=1}^{N}I_{i}\right) =O_{S}\otimes I_{E}  \label{5}
\end{equation}%
The expectation value of these observables in the state $\left\vert \psi
(t)\right\rangle $ is given by%
\begin{equation}
\langle O_{R}\rangle _{\psi (t)}=|a|^{2}\,s_{\Uparrow \Uparrow
}+|b|^{2}\,s_{\Downarrow \Downarrow }+2\func{Re}[ab^{\ast }\,s_{\Downarrow
\Uparrow }\,r(t)]  \label{6}
\end{equation}%
where (see \cite{Max}) 
\begin{equation}
r(t)=\langle \mathcal{E}_{\Downarrow }(t)\left\vert \mathcal{E}_{\Uparrow
}(t)\right\rangle =\prod_{i=1}^{N}\left[ |\alpha
_{i}|^{2}e^{-ig_{i}t}+|\beta _{i}|^{2}e^{ig_{i}t}\right]  \label{7}
\end{equation}%
It is clear that the time-behavior of $\langle O_{R}\rangle _{\psi (t)}$
depends on the behavior of $r(t)$. In order to know such a behavior, we
performed numerical simulations for%
\begin{eqnarray}
|r(t)|^{2} &=&\prod_{i=1}^{N}(|\alpha _{i}|^{4}+|\beta _{i}|^{4}+2|\alpha
_{i}|^{2}|\beta _{i}|^{2}\cos 2g_{i}t)  \notag \\
&=&\prod_{i=1}^{N}f_{i}(t)  \label{8}
\end{eqnarray}%
\newline
where the $\left\vert \alpha _{i}\right\vert ^{2}$\ were obtained from a
random-number generator, and the $\left\vert \beta _{i}\right\vert ^{2}$
were computed as $\left\vert \beta _{i}\right\vert ^{2}=1-\left\vert \alpha
_{i}\right\vert ^{2}$. But before presenting the computer simulations, we
will study the Poincar\'{e} time for this model.

\paragraph{\textbf{Analyzing the Poincar\'{e} time.}}

Each $f_{i}(t)$ of eq.(\ref{8}) comes back to its initial value for the
first time at a time $t_{Pi}$, such that $2g_{i}t_{Pi}=2\pi \Rightarrow
t_{Pi}=\pi /g_{i}$. In turn, $|r(t)|^{2}$ comes back to its initial value
when all the $f_{i}(t)$ do it. Therefore, the Poincar\'{e} time $t_{P}$ of
this model is the time when all the $f_{i}(t)$ come back to their initial
values for the first time. Let us consider three cases:

\begin{enumerate}
\item[(a)] All the $g_{i}$ have the same value: $g_{i}=g$, for all $i$. So,
the mean value is $\overline{g_{i}}=g$. In this case, all the $f_{i}(t)$
come back to their initial values at the same time $t_{Pi}=\pi /g$.
Therefore, the Poincar\'{e} time is $t_{P}=\pi /g=\pi /\overline{g_{i}}$,
and it does not depend on the number of particles $N$.

\item[(b)] All the $g_{i}$ are such that $g_{i}=n_{i}g_{\min }$, with $%
n_{i}\in \mathbb{N}$. In this case, $t_{P}$ is the largest $t_{Pi}$,
corresponding to the smallest $g_{i}$, $g_{\min }$: $t_{P}=\pi /g_{\min }$.
So, given a $g_{\min }$, the Poincar\'{e} time $t_{P}$ does not depend on $N$%
. Since the mean value is $\overline{g_{i}}>g_{\min }$, then $t_{P}>\pi /%
\overline{g_{i}}$.

\item[(c)] All the $g_{i}$ are random. Since in the computations the $g_{i}$
are rational numbers, we can express them as $g_{i}=p_{i}/q_{i}$, \ with $%
p_{i},q_{i}\in \mathbb{N}$. If we make $t_{P}=\pi Q$, the number $Q$ has to
be such that $Q=n_{i}q_{i}/p_{i}$ for all $i$, with $n_{i}\in \mathbb{N}$.
Then, $n_{i}=Q\,p_{i}/q_{i}$. Since the $p_{i}$ and $q_{i}$ are random
natural numbers, the least $Q$ that guarantees that $n_{i}$ is a natural
number for all $i$ is $Q=\tprod\nolimits_{i=1}^{N}q_{i}$. Therefore, 
\begin{equation}
t_{P}=\pi Q=\pi \prod_{i=1}^{N}q_{i}\,  \label{9}
\end{equation}%
In turn, $Q$ is larger than any $q_{i}$, more larger as $N$ increases. Then,
for $N$ large and for any $g_{i}=p_{i}/q_{i}$, the Poincar\'{e} time is $%
t_{P}=\pi Q\gg \pi q_{i}>\pi q_{i}/p_{i}=\pi /g_{i}=t_{Pi}$, and also $%
t_{P}\gg \pi /\overline{g_{i}}$. Moreover, the order to magnitude of $t_{P}$
can be estimated as $t_{P}=\pi Q\sim \pi \overline{q_{i}}^{N}\gg \left( \pi /%
\overline{g_{i}}\right) ^{N}$: when the coupling coefficients are random,
the Poincar\'{e} time increases exponentially with the number of particles.
\end{enumerate}

\paragraph{\textbf{Case 1: Homogeneous environment.}}

Let us consider the case where all the particles of the environment are of
the same kind. In this case, the reasonable assumption is that the particle $%
P$ interacts in the same way with all the environmental particles and, as a
consequence, $g_{i}=g$ for all $i$ (Case (a)). The time-behavior of $%
|r(t)|^{2}$ of eq.(\ref{8}), for $N=100$ and $g_{i}=g=0.5$, is plotted in
Figure 1: we can see that the Poincar\'{e} time $t_{P}=\pi /g=\pi /0.5\simeq
6.28$ is not sufficiently larger than the decoherence time to consider the
result as an effective decoherence that may lead to classicality.

 \begin{figure}[t]

 \centerline{\scalebox{0.7}{\includegraphics{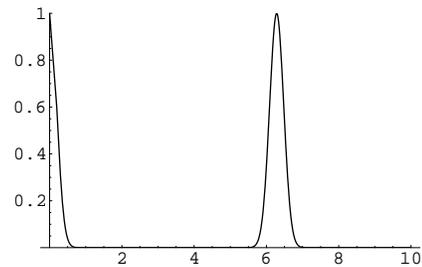}}}
\caption{Plot of $|r(t)|^{2}$ given by
eq.(\protect\ref{8}), for $N=100$ and $g_{i}=g=0.5$. }
 \label{f1}\vspace*{0.cm}
\end{figure}

Now let us consider the case that $P$ does not interact in the same way with
all the environmental particles. In particular, we will suppose that the
coupling coefficients are random in an interval $\left[ \overline{g_{i}}%
-\Delta g,\overline{g_{i}}+\Delta g\right] $ around the mean value $%
\overline{g_{i}}$. The time-behavior of $|r(t)|^{2}$ of eq.(\ref{8}), for $%
N=100$ and $g_{i}$ random, with $\overline{g_{i}}=0.5$ and $\Delta g=0.1$,
is plotted in Figure 2. In this case, the Poincar\'{e} time is much larger
than $\left( \pi /\overline{g_{i}}\right) ^{N}=\left( 6.28\right) ^{100}$
(Case (c)), a value that can be considered infinite for all practical
purposes. Therefore, as Figure 2 shows, it can be legitimately said that the
particle $P$ decoheres and may become classical.

 \begin{figure}[t]

 \centerline{\scalebox{0.7}{\includegraphics{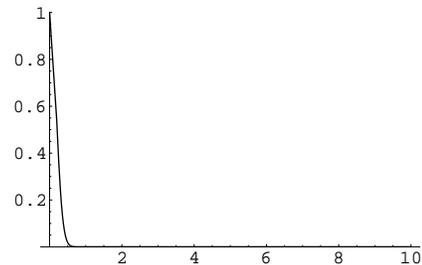}}}
\caption{Plot of $|r(t)|^{2}$ given by
eq.(\protect\ref{8}), for $N=100$ and random $g_{i}\in \left[ 0.4,0.6\right] 
$.}
 \label{f1}\vspace*{0.cm}
\end{figure}

\paragraph{\textbf{Case 2: Non-homogeneous environment.}}

In this case we will consider an environment whose particles are not all of
the same kind, and the particles of each kind $j$ interact with $P$ through
their own coupling coefficient $g_{j}$: $N_{1}$ particles through $g_{1}$, $%
N_{2}$ particles through $g_{2}$, ... and $N_{p}$ particles through $g_{p}$,
such that $\dsum\limits_{j=1}^{p}N_{j}=N$. Then, the $|r(t)|^{2}$ of eq.(\ref%
{8}) can be rewritten as:

\begin{eqnarray}
|r(t)|^{2} &=&\prod_{i=1}^{N}f_{i}(t)=  \label{10} \\
&=&\left( \prod_{i=1}^{N_{1}}f_{i,g_{1}}(t)\right) \left(
\prod_{i=1}^{N_{2}}f_{i,g_{2}}(t)\right) ...\left(
\prod_{i=1}^{N_{p}}f_{i,g_{p}}(t)\right)  \notag
\end{eqnarray}%
where each particular product is the contribution of each kind of particles.
In particular, we will consider a situation where the environment is
composed of $N$ particles that are almost all of the same kind, with the
exception of a slight \textquotedblleft contamination\textquotedblright\ of
particles of different kinds. The time-behavior of $|r(t)|^{2}$ of eq.(\ref%
{10}), for $N=100$, $N_{1}=91$, $N_{2}=N_{3}=N_{4}=3$, $g_{1}=2.4$, $%
g_{2}=1.2$, $g_{3}=0.6$ and $g_{4}=0.3$, is plotted in Figure 3. Since all
the $g_{i}$ are such that $g_{i}=n_{i}g_{\min }=n_{i}$ $g_{4}$, this
situation corresponds to Case (b). Then, the Poincar\'{e} time can be easily
computed as $t_{P}=\pi /g_{\min }=\pi /0.3\simeq 10.43$. Therefore, in
Figure 3, the peak in $10.43$ is the Poincar\'{e} time, but the peaks around
it are \textit{not} due to the recurrence of $|r(t)|^{2}$. This means that
the particle $P$ in interaction with this \textquotedblleft
contaminated\textquotedblright\ environment does not decohere.

 \begin{figure}[t]

 \centerline{\scalebox{0.7}{\includegraphics{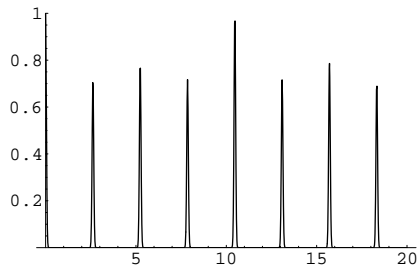}}}
\caption{Plot of $|r(t)|^{2}$ given by
eq.(\protect\ref{10}), for $N=100$, $N_{1}=91$, $N_{2}=N_{3}=N_{4}=3$, $%
g_{1}=2.4$, $g_{2}=1.2$, $g_{3}=0.6$ and $g_{4}=0.3$.}
 \label{f1}\vspace*{0.cm}
\end{figure}

Now we will consider that $P$ does not interact in the same way with the
particles of a same kind $j$, but the coupling coefficients $g_{ji}$ are
random in the intervals $\left[ \overline{g_{ji}}-\Delta g_{j},\overline{%
g_{ji}}+\Delta g_{j}\right] $ around the corresponding mean value $\overline{%
g_{ji}}$: $N_{1}$ particles with $g_{1i}\in \left[ \overline{g_{1i}}-\Delta
g_{1},\overline{g_{1i}}+\Delta g_{1}\right] $, $N_{2}$ particles with $%
g_{2i}\in \left[ \overline{g_{2i}}-\Delta g_{2},\overline{g_{2i}}+\Delta
g_{2}\right] $, ... and $N_{p}$ particles with $g_{pi}\in \left[ \overline{%
g_{pi}}-\Delta g_{p},\overline{g_{pi}}+\Delta g_{p}\right] $, such that $%
\dsum\limits_{j=1}^{p}N_{j}=N$. Then, the $|r(t)|^{2}$ of eq.(\ref{8}) can
be rewritten as:

\begin{eqnarray}
|r(t)|^{2} &=&\prod_{i=1}^{N}f_{i}(t)=  \label{11} \\
&=&\left( \prod_{i=1}^{N_{1}}f_{i,g_{1i}}(t)\right) \left(
\prod_{i=1}^{N_{2}}f_{i,g_{2i}}(t)\right) ...\left(
\prod_{i=1}^{N_{p}}f_{i,g_{pi}}(t)\right)  \notag
\end{eqnarray}%
In particular, we will study a situation similar than the previous one with
respect to the values of $N$ and of the $N_{j}$, and where the $g_{j}$ used
there become here the mean values $\overline{g_{ji}}$. Moreover, in all the
cases the $\Delta g_{j}$ were selected as approximately the 30\% of the
corresponding mean value $\overline{g_{ji}}$. The time-behavior of $%
|r(t)|^{2}$ of eq.(\ref{11}), for $N=100$, $N_{1}=91$, $N_{2}=N_{3}=N_{4}=3$%
, $\overline{g_{1i}}=2.4$, $\overline{g_{2i}}=1.2$, $\overline{g_{3i}}=0.6$, 
$\overline{g_{4i}}=0.3$, $\Delta g_{1}=0.8$, $\Delta g_{2}=0.4$, $\Delta
g_{3}=0.2$ and $\Delta g_{4}=0.1$, is plotted in Figure 4, where we can see
the drastic decoherence of the model. The same result is plotted in Figure 5
with a different time scale, in order to show the decoherence time.

 \begin{figure}[t]

 \centerline{\scalebox{0.7}{\includegraphics{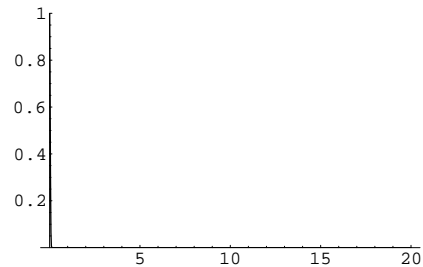}}}
\caption{Plot of $|r(t)|^{2}$ given by
eq.(\protect\ref{11}), for $N=100$, $N_{1}=91$, $N_{2}=N_{3}=N_{4}=3$, and
random $g_{1i}\in \left[ 1.6,3.2\right] $, $g_{2i}\in \left[ 0.8,1.6\right] $%
, $g_{3i}\in \left[ 0.4,0.8\right] $, and $g_{4i}\in \left[ 0.2,0.4\right] $.}
 \label{f1}\vspace*{0.cm}
\end{figure}

 \begin{figure}[t]

 \centerline{\scalebox{0.7}{\includegraphics{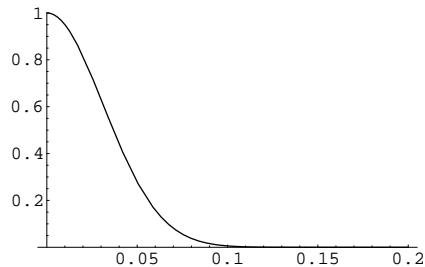}}}
\caption{Same as Fig. 4 with a different
time scale.}
 \label{f1}\vspace*{0.cm}
\end{figure}

\paragraph{\textbf{Conclusions.}}

By means of computer simulations we have proved the great influence that the
randomness of the coupling coefficients exerts on decoherence. Such an
influence appears under two forms: (i) the huge increasing of the Poincar%
\'{e} time, and (ii) the strong damping off of the peaks that preclude
decoherence. Then, in both cases the change from constant coefficients to
random coefficients greatly improves the \textquotedblleft
efficiency\textquotedblright\ of decoherence. It is worth stressing that,
whereas the first effect may be expected, the second effect is not
foreseeable: it is not easy to explain a priori why the randomness of the
coupling coefficients washes off the high peaks present in the non-random
case.

On the other hand, a realistic environment is usually composed by a limited
number of different kinds of particles, and it is reasonable to suppose that
the particle of interest $P$ interacts with all the particles of the same
kind with the same strength. So, the models with random coupling
coefficients are not realistic. As we have pointed out in the Introduction,
this would be not relevant if the randomness of the coefficients had no
significant effect. But now we know that such randomness has a dramatic
influence on decoherence, whether it is expectable or not. Therefore, the
criterion for the selection of the coupling coefficients has to be carefully
analyzed in each model, in order to avoid conclusions drawn from unphysical
results.

\paragraph{\textbf{Acknowledgments.}}

This research was partially supported by grants of FONCYT, CONICET and the
University of Buenos Aires, Argentina.

\end{document}